\documentclass[prb,twocolumn]{revtex4-1} 
\usepackage{amsmath} % needed for \tfrac, \bmatrix, etc.
\usepackage{amsfonts} % needed for bold Greek, Fraktur, and blackboard bold
\usepackage{graphicx} % needed for figures
\usepackage{float}

\newcommand{\be}{\begin{equation}}
\newcommand{\ee}{\end{equation}}
\newcommand{\bea}{\begin{eqnarray}}
\newcommand{\eea}{\end{eqnarray}}
\renewcommand{\ss}[1]{_{\hbox{\tiny #1}}}
\newcommand{\us}[1]{^{\hbox{\tiny #1}}}

\newcommand{\comment}[1]{}

\begin{document}

\title{Learning universality and scaling from simple deposition models}
% In a long title you can use \\ to force a line break at a certain location.

\author{Alessandro Santini}
\email{alessandro.santini2@stud.unifi.it} % optional
% If there were a second author at the same address, we would put another 
% \author{} statement here. Don't combine multiple authors in a single
% \author statement.
\affiliation{Dipartimento di Fisica e Astronomia, Universit\`a di Firenze,
via G. Sansone 1, 50019 Sesto Fiorentino, Italy}
% Please provide a full mailing address here.

\author{Paolo Politi}
\email{paolo.politi@cnr.it}
\affiliation{Istituto dei Sistemi Complessi, Consiglio Nazionale delle Ricerche,
via Madonna del Piano 10, 50019 Sesto Fiorentino, Italy}
\affiliation{Istituto Nazionale di Fisica Nucleare, Sezione di Firenze,
via G.\ Sansone 1, I-50019 Sesto Fiorentino, Italy}

\date{\today}

\begin{abstract}
We use deposition models of kinetic roughening
of a growing surface to introduce the concepts of universality and 
scaling and to analyze the qualitative and quantitative role of different parameters.
In particular, we focus on two classes of models where the deposition is accompanied by a local
relaxation process within a distance $\delta$. The models are in the Edwards-Wilkinson
universality class, but the role of $\delta$ is nontrivial.
\end{abstract}

\maketitle % title page is now complete

\section{Introduction} 

Kinetic roughening is a widespread phenomenon that is related to the
dynamics of a driven interface in the presence of noise. The latter makes
the interface rough; that is, the mean square deviation of the interfacial position
diverges at large temporal and spatial scales.
Examples of driven interfaces include flame fronts, wetting fronts,
magnetic domain walls, and the growing front of a bacterial colony.\cite{barabasi, meakin, growth}
In this paper we study the dynamics of the interface between 
a solid and a vacuum phase when particles are deposited ballistically
onto a growing surface. Ballistic deposition means that particles arrive
following a straight trajectory, normal to the average orientation of the surface to avoid shadowing effects. 
This situation is qualitatively different
from a solid phase that is in contact with a gas phase, because in the latter case
particles undergo a diffusion process, which gives rise to a nonlocal
growth process whose morphology resembles diffusion limited aggregation.\cite{Sander}

To be specific and to keep the notation simple, we consider
a one-dimensional discrete interface defined by the local height $h_i$, 
where $i=1,\dots, L$ and $h_i\ge 0$ are integers. 
The roughness $W(L,t)$,
which depends on the time $t$ and the size $L$ of the growing interface,
is defined as the square root of the variance of the height,
\be
W^2(L,t) = \left\langle \overline{h^2} - {\overline{h}}^2 \right\rangle,
\label{W}
\ee
where the horizontal bar means a spatial average,
$\overline A = (1/L)\sum_{i=1}^L A_i$, and the brackets 
$\langle \cdots\rangle$ denote the average over noise.

The models of deposition plus relaxation that we will consider can be
summarized by the following algorithm:
\begin{enumerate}
\item Choose a random deposition site $k$ between 1 and $L$.
\item Apply a relaxation rule to choose the incorporation
site $j$ in the interval $[k-\delta,k+\delta]$. Use periodic boundary 
conditions such that $h_{j+L}=h_j$.
\item Increment the height of the incorporation site,
$h_j \to h_j +1$.
\end{enumerate}
The models to be discussed differ in the relaxation rule (2)
and the value of the interval $\delta$. 
A random integer $m$ among $M$ values, $m=1,2,\dots,M$, is obtained from a random number generator that
returns a random real number $x$ in the interval $[0,1)$ by the relation
$m=\texttt{INT}(xM)+1$, where $\texttt{INT}$ means the integer part.

In a deposition model, there are deterministic and random ingredients.
In this paper the initial configuration is always a flat surface of zero height,
$h_i(t=0) =0$ $\forall i$. Alternatively, we can use an initial surface with some roughness,
which could be deterministic (such as a pyramidal profile) or a surface with
some degree of randomness. The asymptotic form of $W(L,t)$ is independent 
of the initial configuration (see the Appendix).
It is straightforward to see that step (1) is random and step (3) is deterministic; step (2) is mainly deterministic, but with a random component that is
necessary (as we will see) to break a tie without breaking a symmetry.

\section{Models and scaling function}
\label{sec.scaling}

\begin{figure}[h!]
\includegraphics[scale=0.2]{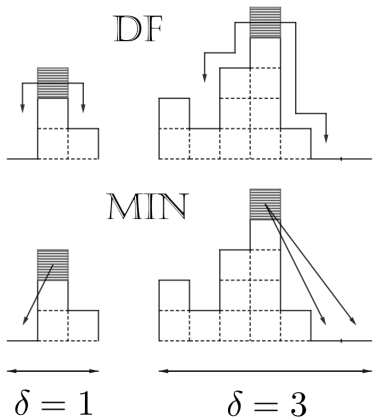}
\caption{\label{etichetta}Schematic diagrams for the particle movements in the downward funneling and the minimum models with $\delta=1$ and $\delta=3$. 
Gray particles are newly deposited particles. If there is more than one arrow from the deposition particle, a random choice
from the incorporation sites is made.}
\end{figure}

We will consider a downward funnelling model and a minimum model,\cite{dep_mod}
as illustrated in Fig.~\ref{etichetta}. In the downward funnelling model (DFM) the deposited particle may reduce its height by moving to a
neighboring site for a maximum of $\delta$ hops. If both neighbors of the deposition site
have a lower height, a random choice is made.
In the minimum model (MIN), the particle is incorporated at the site of minimum 
height within the interval $[k-\delta,k+\delta]$. For equivalent minima,
one site is chosen at random. We see that step (2) is deterministic
and randomness is invoked when there are equivalent incorporation sites.

\begin{figure}[h!]
\includegraphics[scale=0.34]{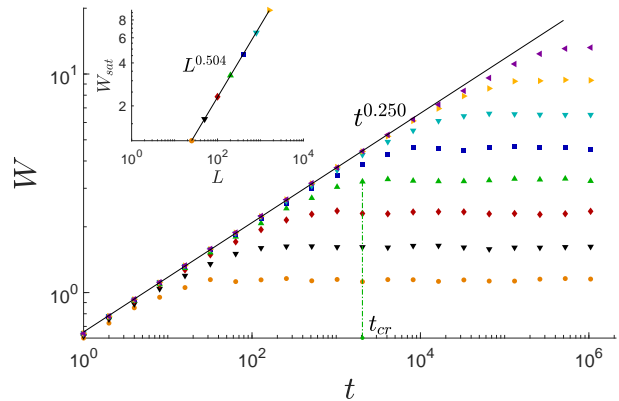}
\caption{\label{DF1}The time dependence of $W(L,t)$ for $L=25$, 50, 100, 200, 400, 800, 1600, and 3200 for the downward funnelling model with $\delta=1$. Increasing values of $L$ correspond
to increasing values of $W(L,t=\infty)$. The slope of the solid line is $\beta=0.25$. 
The inset shows the $L$ dependence of the stationary values of the roughness $W_{\text{sat}}$. The slope of the solid line is $\alpha=0.504$.
The quantity $t\ss{cr}$ is the crossover time between the power-law regime, $W\simeq t^\beta$, and the stationary regime,
$W\simeq L^\alpha$.}
\end{figure}

We highlight in Fig.~\ref{DF1} several general features of the roughness for the downward funnelling model with $\delta=1$. For fixed $L$, there is an initial temporal regime during which $W(L,t)$ increases with time $t<t\ss{cr}$, followed by a second, stationary regime for $t>t\ss{cr}$ during which $W(L,t)$ saturates. The crossover time $t\ss{cr}$ increases with $L$, and the
saturation value increases as well.
The log-log plots in Fig.~\ref{DF1} suggest two power laws,
$W(L\to\infty,t) \simeq t^\beta$ and $W(L,t\to\infty) \simeq L^\alpha$, where $\alpha$ is the roughness exponent and $\beta$ is
the growth exponent. The crossover time can be derived from
the relation $t\ss{cr}^\beta \simeq L^\alpha$; that is,
$t\ss{cr} \simeq L^{\alpha/\beta} \equiv L^z$, which defines the dynamical
exponent $z=\alpha/\beta$.
From the simulation results we conclude that
\be
\alpha\ss{DF1} \simeq \frac{1}{2}, \quad
\beta\ss{DF1} \simeq \frac{1}{4}, \quad
z\ss{DF1} \simeq 2,
\ee
where the subscript DF1 denotes the downward funnelling model with $\delta=1$.

\begin{figure}[h!]
\includegraphics[scale=0.2575]{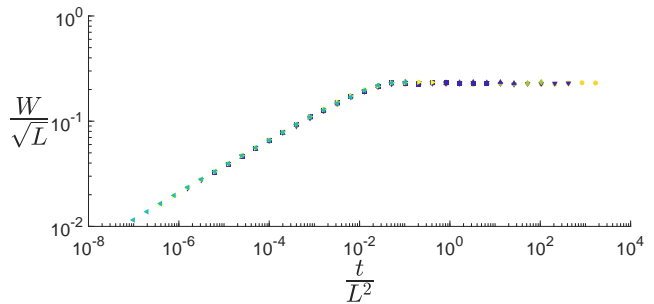}
\caption{Plot of $W/L^{\alpha}$ against $t/L^{z}$ for the downward funnelling model with $\delta=1$ showing data collapse.}\label{DF1scaling}
\end{figure}

If we rescale time with respect to the crossover time and rescale the roughness
with respect to its saturation value, we obtain data collapse
because curves for different values of $L$ now lie on the same curve as shown in Fig.~\ref{DF1scaling}. These results suggest the 
following scaling form for the roughness,\cite{FV,FVbook,JK}
\be
W(L,t) = L^\alpha w(t/L^z)
\label{scaling1}
\ee 
with
\be
w(u) \simeq \left\{ 
\begin{array}{cc}
u^\beta & u \ll 1 \\
\mbox{const.} & u \gg 1
\end{array}
\right.
\, .
\label{scaling2}
\ee
The relation \eqref{scaling1} means that if we rescale space by the factor $b$, it is possible to
rescale time and the height of the interface so as to preserve Eq.~(\ref{scaling1}).
To see this scaling note that if $L\to bL$, the constancy of the argument of the function $w(u)$ 
(necessary because $w$ is not an homogeneous function)
requires that $t\to b^z t$, and $W$ acquires the factor $b^\alpha$. 
According to the definition of the roughness in Eq.~(\ref{W}), this factor is equivalent
to rescaling the height as $h\to b^\alpha h$.

The growing interface is therefore self-similar so that if we do a suitable magnification of a
snapshot of the surface in the stationary regime $t\gg t\ss{cr}$, the enlarged image is statistically equivalent to the interface itself. Otherwise, a suitable magnification of the snapshot is statistically equivalent to the interface at a different time.
The only condition is that we must consider large temporal and spatial scales, $t,L\to\infty$, in the same way that scaling appears
near the ferro--paramagnetic phase transition in the Ising model.~\cite{scaling,widom}

\section{Universality}
\label{sec.universality}

In analogy with continuous thermal phase transitions, we expect that certain features
of the deposition model are irrelevant; that is, the roughness exponents and the function
$w(u)$ do not change. To check this expectation,
we plot $W/L^{1/2}$ as a function of $t/L^2$ for 
DF1, DF10, MIN1, and MIN2 (DF10 refers to the DF model with $\delta = 10$, and similar notation for the MIN models).

\begin{figure}[h!]
\includegraphics[scale=0.2575]{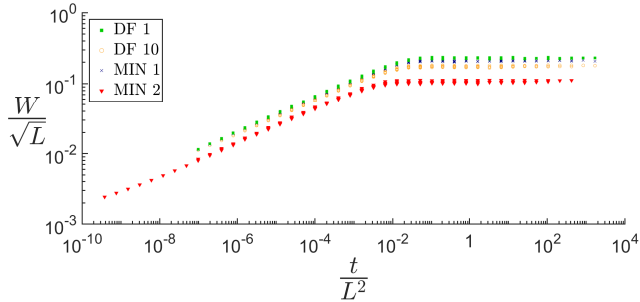}
\caption{The scaled roughness ($W/\sqrt{L}$) as a function of the scaled time ($t/L^2$) for different models.
Data collapse for the same model is related to scaling as discussed in Sec.~\ref{sec.scaling}. The fact that curves for different models 
require an additional scaling of the axes by a numerical factor to superpose is related to universality (see Sec.~\ref{sec.universality}),
and to nonuniversal features (see Sec.~\ref{sec.nonuniversality}).}
\label{collapse1}
\end{figure}

The results show that the DF$\delta$ and MIN$\delta$ models have the same exponents, because
after rescaling the four models we have considered, they exhibit the same initial slope (therefore
the same exponent $\beta$) and $W/\sqrt{L}$ asymptotically goes to a constant and therefore they have the same exponent $\alpha=1/2$.
The residual differences among models are non-universal features, which we will discuss in Sec.~\ref{sec.nonuniversality}.

The models we have introduced belong to the same Edwards-Wilkinson (EW) universality class.\cite{EW,JK,libro} 
Its large scale properties are described by the
homogeneous, linear stochastic differential equation\cite{EW}
\be
\partial_t h(x,t) = \nu \partial_{xx} h(x,t) + \eta(x,t),
\label{EW}
\ee
where $\eta(x,t)$ indicates white noise whose properties
can be formally written as
\begin{subequations}
\label{noise}
\begin{align} 
\langle \eta(x,t)\rangle &=0 \\
\langle \eta(x,t)\eta(x',t')\rangle &=\Gamma\delta(x-x')\delta(t-t').
\end{align}
\end{subequations}

In the Appendix we give a short analytical derivation of the scaling function and of
the roughness exponents for the Edwards-Wilkinson equation, which are equal to
\be
\alpha\ss{EW} = \frac{1}{2}, \quad
\beta\ss{EW} = \frac{1}{4}, \quad
z\ss{EW} = 2.
\ee

Rather than discussing the exact form of the scaling function, we will consider
the Edwards-Wilkinson equation itself and discuss which features are relevant or irrelevant.
We first note that the continuum height $h(x,t)$ in
Eq.~(\ref{EW}) can be interpreted as the local height with respect to the average height
of the interface. A slightly more general version of Eq.~(\ref{EW}) can be
obtained by adding a constant term $F$ to the right-hand side, representing the average
flux of particles arriving at the surface, while $\eta(x,t)$ represents its fluctuating part.
By redefining $z(x,t) = h(x,t) - Ft$ we can eliminate the $F$ term, which is the reason why
it is common not to include it.

A second, non-trivial remark is that Eq.~(\ref{EW}) has a conserved form,
because $\nu \partial_{xx} h = -\partial_x J$ with $J= -\nu \partial_x h$, and has up-down symmetry, because $\tilde h(x,t)=-h(x,t)$ satisfies 
the same Edwards-Wilkinson equation.\cite{note1}
These two properties are essential for determining what deposition processes belong to the Edwards-Wilkinson universality classes: they must conserve matter and volume,
and the up-down symmetry must be conserved. These properties are worthy of
a short comment. 

Matter conservation is not broken if, for example, each surface particle has a constant rate of desorption,
because such a process would simply lead to a redefinition of the flux $F$, and therefore of the
average growth $Ft$. Instead, matter conservation is broken if the deposition process 
is accepted or rejected according to the local configuration of the interface. A simple way to
implement non-conservation is to force the restricted solid-on-solid rule,\cite{BD-RSOS,RSOS}
according to which
$|h_{i+1} - h_i| \le 1$ at all times. If the deposition of a particle breaks such a rule,
a new deposition site is randomly chosen. 
We can also have matter conservation without having volume conservation because voids or overhangs form
during the growth process. This [lack of volume conservation is the case for the ballistic deposition model,\cite{BD-RSOS}
where a deposited particle (traveling ballistically in the vertical direction) 
stops as soon as it meets the growing interface. In this case
the evolution rules (2) and (3) can be replaced by
\begin{equation}
h_k \to \max(h_{k-1},h_k +1,h_{k+1}).
\end{equation}
Both non-conserving models (restricted solid-on-solid and ballistic deposition)
belong to a very important, nonlinear universality class called the Kardar-Parisi-Zhang 
universality class,\cite{KPZ} whose continuum description corresponds to the Edwards-Wilkinson equation with the
addition of a nonconserved
term on the right-hand side, $\lambda (\partial_x h)^2$, which also breaks the up-down symmetry

It might be expected that {\it any} deposition model breaks the up-down symmetry,
but that is not so.
If $h_i$ is interpreted as the number of particles in the $i$th box rather than as a height and the
relaxation process is thought of as the transfer of particles from a box with many particles to a neighboring box with fewer particles,
then no up-down symmetry is broken and the transfer process is immediately recognized to
be a diffusion process, and it is not surprising that it is described by a continuum
diffusion equation.

A final remark about universality concerns the spatial dimension $d$.
We have discussed the deposition models with relaxation in one spatial dimension
not (only) because simulations are simpler, but because $d=1$ is the sole integer dimension that
provides the simple scaling picture we have described.
In the theory of equilibrium phase transitions a given universality class will have a lower ($d_\ell$) and an upper
($d_u$) critical dimension. For $d <d_\ell$ there is no phase transition and for $d>d_u$ systems are described 
by mean-field models.\cite{critical_dimension} 
Fluctuations decrease in importance as the spatial dimension is increased,
and thus the surface becomes less rough with increasing $d$,
becoming non-rough if $d>d_u$ (non-rough means that $\lim_{t,L\to\infty} W(L,t)$ is finite).
The extension~\cite{JK,libro} to general $d$ of the analytical derivation in the Appendix shows that
$d\us{EW}_\ell=0$ and $d\us{EW}_{u}=2$. For $d=2$, $W(L,t)$ increases logarithmically rather than as a power law,
and for $d < d_\ell$ the roughness diverges at increasing times even for finite $L$ due to a divergence
of the local slope of the interface.

\section{Non-universal parameters}
\label{sec.nonuniversality}

\begin{figure}[h]
\includegraphics[scale=0.32]{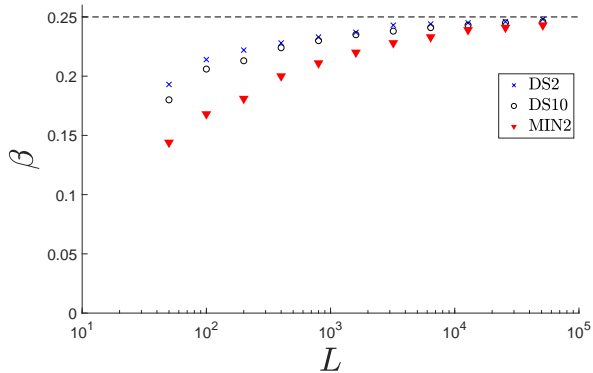}
\caption{The $L$-dependence of the exponent $\beta$ for the models, DF2, DF10 and MIN2. The universal value of $\beta$ is well defined for large $L$. Due to finite size effects, $\beta < 0.25$ for small $L$. 
Note that finite size effects are more important for MIN2.
%{\bf Fare simboli diversi.}
}\label{betaplot}
\end{figure}

We now discuss the effect of changing
an irrelevant parameter ($\delta$ in our case) or modifying the relaxation rule
while remaining in the same universality class (downward funnelling rather than MIN). Figures~\ref{collapse1} and \ref{betaplot} can be the starting point of this discussion,
because they lead to qualitative and quantitative conclusions.
The qualitative conclusion is that the relaxation process is more effective
(that is, $W$ is lower) if $\delta$ is larger or if we go from the downward funnelling model to the minimum model.
The quantitative conclusions are twofold: changing $\delta$ for the downward funnelling model 
by a factor of ten has less effect that changing $\delta$ by a factor of two
for the minimum model; also the MIN2 model has important finite size effects (see Fig.~\ref{betaplot})
that force us to investigate larger temporal and spatial scales to attain the scaling regime.

\begin{figure}[h]
\includegraphics[scale=0.27]{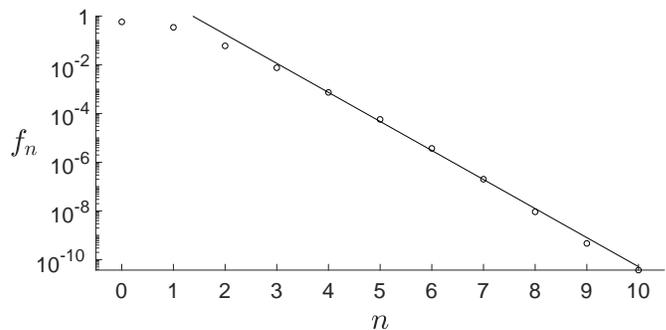}
\caption{The discrete probability distribution of downward hops of size $n=\lvert j-k \rvert$. The solid line represents the exponential decay of the distribution $ f_n \approx e^{-n/3}$.}
\label{istogramma}
\end{figure}

Finite size effects are not so important for the DF$\delta$ models
where the model DF10 is only slightly different from the DF1 model,
and it is practically equivalent to the model DFL ($\delta=L$).
This similarity among DF$\delta$ models can be understood from Fig.~\ref{istogramma} which shows
the frequency of downward hops of size $n=|j-k|$ for the DFL model.\cite{note2}
According to Fig.~\ref{istogramma}, the frequency decreases exponentially, $f_n \approx e^{-n/3}$,
making irrelevant the events for which $n$ 
is significantly larger than $n_0=3$. Therefore, all DF$\delta$ models with 
$\delta > n_0$ are practically equivalent.

The minimum model has a completely different behavior with increasing $\delta$
because the deposited particles search for the lowest height site within a distance
$\delta$ and the surface profile between the deposition site $k$ and the
incorporation site $j$ is irrelevant. 
For any finite $\delta$ (more precisely, for any value of $\delta$ that does not scale
with $L$), the minimum model belongs to the same Edwards-Wilkinson universality class, but simulations require increasingly larger $t$ and $L$ (and are therefore more computationally demanding) with increasing $\delta$.

We now discuss the effect of the irrelevant quantities ($\delta$ and the relaxation criterion)
on the scaling function $w(u)$. The analytical expression for $w(u)$ (see the Appendix) is 
\begin{subequations}
\label{ScalingFunction}
\begin{align}
W(L,t)&=\sqrt{\frac{\Gamma L}{2 \nu}}\; w\left(\frac{\nu t}{L^2}\right) \\
w^2(u)&=\frac{1}{\pi}\!\int\limits_{\pi}^{\infty} ds\, \frac{1-e^{-2s^2u}}{s^2}.
\end{align}
\end{subequations}

\begin{figure} [h]
\includegraphics[scale=0.25]{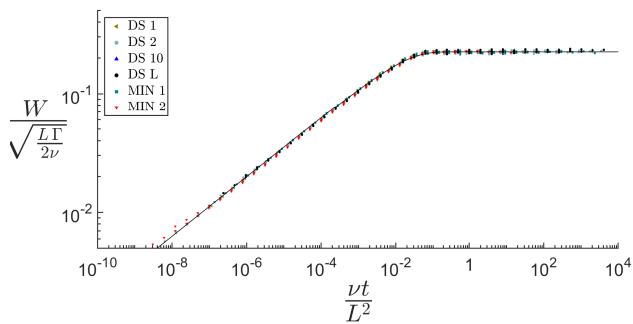}
\caption{Plot of the data collapse for different models. The solid line represents $w(\nu t/L)$. }
\label{collapse2}
\end{figure}

By using the analytical expression for $W(L,t)$ and comparing it to the numerical profiles for the roughness
obtained by the simulations, it is possible to extract $\nu$ and $\Gamma$ for each model\cite{Vvedensky}
(see Table~I). We then use these values to rescale the axes and obtain a single scaling curve
for {\it all} models belonging to the EW universality class (see Fig.~\ref{collapse2}).

The noise appears to depend very weakly on the details of the relaxation process because its main source is
the noise of the deposition process, which is the same in all simulations.
The variability of $\mu$ reflects what we anticipated: relaxation is more effective for the MIN
model and within the same model is more effective with increasing $\delta$.
As anticipated, 
\be
\frac{\nu\ss{MIN}(\delta=2)}{\nu\ss{MIN}(\delta=1)} > \frac{\nu\ss{DF}(\delta=10)}{\nu\ss{DF}(\delta=1)} .
\ee
Finally, we expect that $\nu\ss{DF}(\delta=L) \simeq \nu\ss{DF}(\delta=10) \simeq 2.4$.
As mentioned,
if we use the numerical values of $\nu$ and $\Gamma$ from Table~1 to rescale the roughness,
we obtain a single curve for all models as shown in Fig.~\ref{collapse2}.

\begin{table}[h]
\vspace{0.2cm}
\begin{minipage}{.48\linewidth}
{\large\begin{tabular}{c|c|c}
\hline
\multicolumn{3}{c}{DF} \\ \hline
$\delta$ & $\nu$ & $\Gamma$\\
\hline
\ \,$1$ & $1.2$ & $1.2$ \\
\ \,$2$ & $2.0$ & $1.3$ \\
$10$ & $2.4$ & $1.4$ \\
\hline
\end{tabular}}
\end{minipage}\quad
\begin{minipage}{.48\linewidth}
\vspace*{-0.55cm}
{\large\begin{tabular}{c|c|c}
\hline
\multicolumn{3}{c}{MIN} \\ \hline
$\delta$ & $\nu$ & $\Gamma$\\
\hline
\ \,$1$ & $1.4$ & $1.2$ \\
\ \,$2$ & $6.1$ & $1.3$ \\
\hline
\end{tabular}}
\end{minipage}
\caption{Estimated values of $\Gamma$ and $\nu$ for different models. 
The values are estimated by superposing the curve of Eq.~\eqref{ScalingFunction} on the data collapse of Fig.~\ref{collapse1}. }
\end{table}

\begin{figure}[h]
\includegraphics[scale=0.25]{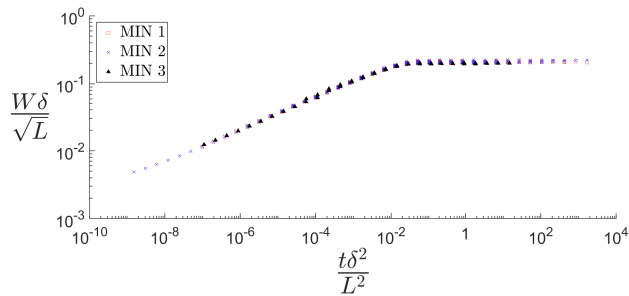}
\caption{The roughness of the minimum models scaled with respect to $L$ and with respect to $\delta$.
The data collapse suggests that $\nu\ss{MIN}(\delta) = \nu_0 \delta^2$.}
\label{scaling_delta}
\end{figure}

We see that $\Gamma$ is essentially independent of $\delta$
in both models and $\nu\ss{DF}$ is almost constant for $\delta > n_0=3$. 
The last question is how $\nu\ss{MIN}$ depends on $\delta$.
According to Table~I, $\nu\ss{MIN}(\delta=2)/\nu\ss{MIN}(\delta=1) \approx 4$,
which suggests the possible scaling $\nu\ss{MIN}(\delta) = \nu_0 \delta^2$.
In Fig.~\ref{scaling_delta} we test this assumption and find a good collapse
for the curves at $\delta=1$, 2, and 3.

\section{Conclusions}

Our main goal has been to suggest that the kinetic roughening of deposition models
can be studied by students with a limited physical and mathematical
background, and that this topic can be a novel
way to approach scaling and universality.
We have also discussed how
two relaxation models depend on $\delta$, showing in particular that
\begin{enumerate}
\item The minimum model has important finite size effects that are already visible for $\delta=2$. These finite size effects are due to the fact that $\nu\ss{MIN}(\delta)$
increases quadratically with $\delta$, see Fig.~\ref{scaling_delta}.

\item The downward funnelling model is well defined for increasing $\delta$, because the probability $f_n$
of $n$ hops decays exponentially, as seen in Fig.~\ref{istogramma}.
\end{enumerate}

We conclude by suggesting further directions that
readers might desire to investigate.

\begin{enumerate}

\item We propose to investigate the kinetic roughening process at
the upper critical dimension of the Edwards-Wilkinson universality class, $d=d\us{EW}_u=2$.
In this case the roughness increases very slowly (logarithmically)
and it might be necessary to investigate unusually large temporal and spatial scales to attain the scaling
regime (we have not checked how large). 
We suggest using the rule with the least relaxation, that is, the DF1 model and 
with an initially flat surface, as in the simulations we have discussed.
The statistical average over noise does not require special attention.
%Plot the roughness on a linear scale and time on a log scale. 

It may be of interest to point out a different ``deposition with relaxation" model whose upper
critical dimension is $d_u=1$.~\cite{JK,KM}
The evolution of the surface is described by Eq.~\ref{h(q,t)} with $\omega_q = \nu |q|$
and scaling in $d=1$ is shown in Fig.~1 of Ref.~[\onlinecite{KM}]. 
Note the log-lin scale of the figure, which should also be used when
studying the DF1 model for $d=2$.

\item In addition to the roughness, which is an average over noise, it would be
interesting to study the distribution of the random variable $h_i(t) - \overline h_i (t)$.
In a continuum picture we can replace Eq.~(\ref{hq}) by Eq.~(\ref{aF}),
obtaining
\be
h(x,t) = \frac{1}{2\pi} \!\int dq e^{iqx} \!\!\int_0^t dt' \eta(q,t') e^{-\omega_q (t-t')},
\ee
which shows that the height $h(x,t)$ is a linear combination of independent random variables
$\eta(q,t')$. According to the central limit theorem 
the distribution follows a Gaussian distribution with zero mean and standard deviation equal to the
roughness $W(L,t)$. Is it possible to find numerically the distribution of $h(x,t) -\overline h(x,t)$ (at fixed $t$) 
for the discrete model?

\item Consider a deposition model for which right-left ($x\to -x$) symmetry has been broken,
by introducing a probability $p\ne 1/2$ in the DF$\delta$ models.
If the relaxation process consists of reducing the height by hopping either to the right or to the left,
we can break right-left symmetry by moving to the right with probability $p$ and to the left with probability $1-p$. 
Modify the EW equation by adding the simplest term that breaks the $x\to -x$ symmetry without
breaking all the other symmetries: up-down symmetry ($h\to -h$) and translation invariance in space ($x\to x + x_0$)
and time ($t\to t+t_0$).
Show that a simple Galilean transformation leads back to the EW equation,
without modifying the roughness exponent. Check this theoretical result numerically
for the most asymmetric case ($p=0$ or $p=1$).

\end{enumerate}

\appendix*

\section{The Edwards-Wilkinson equation}
In the following we sketch the derivation of the Edwards-Wilkinson equation for $d=1$.
For more details and for $d>1$,
see Ref.~\onlinecite{libro}, Sec.~5.6.2.

Because of its linearity we can rewrite the Edwards-Wilkinson equation~(\ref{EW}) as
\be
\partial_t h(x,t) = \nu \partial_{xx} h(x,t) + \eta(x,t),
\ee 
We Fourier transform both sides and obtain
\be
\label{thiseq}
\partial_t h(q,t) = - \nu q^2 h(q,t) + \eta(q,t).
\ee
Equation~\eqref{thiseq} can be solved by multiplying both sides by $e^{\omega_q t}$ and then integrating:
\be \label{h(q,t)}
h(q,t) = h(q,0)e^{-\omega_q t} + e^{-\omega_q t} \!\int_0^t\,dt' \eta(q,t') e^{\omega_q t'},
\ee
where $\omega_q = \nu q^2$.
The first term on the right-hand side is related to the interface profile at $t=0$ and is negligible for large times. 
For simplicity we will assume an initial perfectly flat interface, $h(q,0)=0$, so that
\be
h(q,t) = \!\int_0^t\,dt' \eta(q,t') e^{-\omega_q (t-t')}.
\label{hq}
\ee
By using the correlation of noise in Fourier space,
\begin{subequations}
\begin{align}
\langle \eta(q,t') \eta(q',t'') \rangle & =
\!\int\! dx \!\int\! dx' e^{-iq\cdot{x}} e^{-i{q'}\cdot{x'}}
\langle \eta({x},t') \eta({x'},t'') \rangle \\
& = 2\pi \Gamma \delta(t'-t'') \delta({q}+{q'}) ,
\end{align}
\end{subequations}
we find 
\be
\langle h({q},t) h({q'},t) \rangle =
2\pi \Gamma \left(
\frac{1-e^{-2\omega_{q} t}}{2\omega_{q}} \right) \delta({q}+{q'}) .
\ee
We then use the inverse Fourier transform,
\be
h({x},t) = \frac{1}{2\pi} \!\int\! d{q} e^{i{q}{x}} h({q},t) ,
\label{aF}
\ee
to find that the roughness becomes
\begin{align}
W^2(L,t) &= \frac{1}{(2\pi)^{2}} \!
\int d{q} \! \int d{q'} e^{i({q}+{q'}){x}}
\langle h({q},t) h({q'},t) \rangle\\
&=\frac{\Gamma}{\pi} \!\int_{\pi/L}^{\infty} 
dq \frac{1-e^{-2\omega_{q} t}}{2\omega_{q}} .
\end{align}
We make the change of variables $s=Lq$ and find 
\begin{align}
W(L,t) & =\sqrt{\frac{\Gamma L}{2 \nu}}\; w\left(\frac{\nu t}{L^2}\right) \\
\noalign{\noindent where}
w^2(u)&=\frac{1}{\pi}\!\int_{\pi}^{\infty} ds\, \frac{1-e^{-2s^2u}}{s^2}.
\end{align}

%\begin{acknowledgments} \end{acknowledgments}


\begin{thebibliography}{99}
\bibitem{note2}Recall that $k$ is the deposition site and $j$ is the incorporation site.
In the downward funnelling model the height of the interface decreases
at each lattice site as $|j-k|$ is increased.
\end{thebibliography}
\end{document}